\documentclass{aa}
\usepackage{graphicx,longtable,lscape}
\usepackage{txfonts,psfig}
\usepackage[]{natbib}
\def \ferg {erg cm$^{-2}$ s$^{-1}$}
\def \hcm {\hbox {\ifmmode $ atom cm$^{-2}\else atom cm$^{-2}$\fi}}

\begin{document}
   \title{The Palermo {\it Swift}-BAT Hard X-ray Catalogue\\} 
   \subtitle{II. Results after 39 months of sky survey\\}
   \author{G.\ Cusumano\inst{1}, V.\ La Parola\inst{1}, 
           A.\ Segreto\inst{1}, V.\ Mangano\inst{1},  C.\ Ferrigno\inst{1,2,3}, A.\ Maselli\inst{1},
	   P.\ Romano\inst{1}, T. Mineo\inst{1}, B.\ Sbarufatti\inst{1}, S. Campana\inst{4}, 
	   G. Chincarini\inst{5,4}, P. Giommi\inst{6}, N. Masetti\inst{7}, A. Moretti\inst{4},
           G. Tagliaferri\inst{4}}

   \offprints{G. Cusumano, cusumano@ifc.inaf.it}
   \institute{INAF, Istituto di Astrofisica Spaziale e Fisica Cosmica di Palermo, 
	Via U.\ La Malfa 153, I-90146 Palermo, Italy 
      \and
          Institut f\"ur Astronomie und Astrophysik T\"ubingen (IAAT)
      \and
          ISDC Data Centre for Astrophysics,
          Chemin d'\'Ecogia 16,
          CH-1290 Versoix,
          Switzerland
      \and
	  INAF -- Osservatorio Astronomico di Brera, Via Bianchi 46, 23807 Merate, Italy
      \and
	  Universit\`a degli studi di Milano-Bicocca, Dipartimento di Fisica, Piazza delle Scienze 3, I-20126 Milan, Italy
      \and
	  ASI Science Data Center, via Galileo Galilei, 00044 Frascati, Italy
      \and
          INAF, Istituto di Astrofisica Spaziale e Fisica Cosmica di Bologna, 
        via Gobetti 101, I-40129 Bologna, Italy
     	  }
            
   \date{}
   %\date{Received 1 March 2007/Accepted 4 April}
\abstract
{}
{We present the Palermo {\it Swift}-BAT hard X-ray catalogue obtained from the
analysis of the the data relative to the first 39 months of the Swift mission.
}
{We have developed a dedicated software to perform data reduction, mosaicking
and source detection on the BAT survey data. We analyzed the BAT dataset in 
three energy bands (14--150 keV, 14--30 keV, 14--70 keV), obtaining a list of 
962 detections above a significance threshold of 4.8 standard deviations. The
identification of the source counterparts was pursued using three strategies:
cross-correlation with published hard X-ray catalogues, analysis of field
observations of soft X-ray instruments, cross-correlation with the SIMBAD
databases.
}
{The survey covers 90\% of the sky down to a flux limit of 
$ 2.5\times10^{-11}$\ferg~  and 50\% of the sky down 
to a flux limit of 
$1.8\times10^{-11}$\ferg~ in the 14--150 keV band. We derived a catalogue of 
754 identified sources, of which  $\sim69$\% are extragalactic, 
$\sim27$\% are Galactic objects, $\sim4$\% are already 
known X-ray or gamma ray emitters whose nature has not been determined yet. 
The integrated flux of the extragalactic 
sample is $\sim1\%$ of the Cosmic X-ray background in the 14--150 keV range.
}
{}
\authorrunning {G.\ Cusumano et al.}
\titlerunning {The Swift/BAT Survey. II. }
\maketitle

\section{Introduction\label{intro} }
The study of Galactic and extragalactic sources at energies
greater than 10 keV is fundamental to investigate non thermal
emission processes and to study source populations that are not
detectable in the soft X-ray energy band because their emission
is strongly absorbed by a thick column of gas or dust. Another
major aim of deep and sensitive surveys in the hard X-ray domain
is to resolve the diffuse X-ray background (CXB) and
identify which class of sources gives the larger contribution:
while the CXB at energies lower than 10 keV has been almost
entirely resolved (80--90\%, \citealp{moretti03,worsley05,worsley06,brandt05}),
only a $\sim$1.5\% of the
CXB at higher energies can be associated with resolved sources
\citep{ajello3}

Up to now, the observation of the hard X-ray sky has not been performed with 
%The hard X-ray sky is not accessible to
imaging grazing incidence telescopes because
the reflectivity above 10 keV rapidly falls down due to the
steep decrease of the critical angle with energy. The first surveys in 
the hard X-ray domain were performed with detectors equipped
with collimator-limited field of view: UHURU  (2--20 keV; \citealp{forman78})
and HEAO1 (0.2 keV -- 10 MeV; \citealp{wood84}).
Later, sky images for energies greater than 10 keV have been produced using
coded mask detectors (e.g. \citealp{fenimore78}; \citealp{skinner87a}): 
in such detectors the entrance window of the telescope is partially masked and 
the ``shadows'' of the cosmic sources are projected onto a position-sensitive 
detector. Dedicated algorithms are then used to reconstruct the
position and intensity of the sources in the field of view and, therefore, 
reproduce the image of the observed sky.
In the last two decades space observatories equipped with this type of
telescopes have surveyed the sky reporting detections of numerous sources
emitting in the hard X-ray domain: Spacelab/XRT \citep{skinner87b},
MIR/KVANT/TTM \citep{sun91}, GRANAT/ART-P \citep{pavlinsky92,pavlinsky94},
GRANAT/SIGMA \citep{cordier91,sun91} and
{\it BeppoSAX}/WFC \citep{jager97}.
Today, the IBIS-ISGRI camera \citep{ibis,isgri} on the INTEGRAL observatory 
\citep{integral} with its field of view of $8^{\circ}\times8^{\circ}$ (fully 
coded) is carrying out a 
hard X-ray survey focussing mostly on the Galactic plane in the 20--150 keV 
energy band  with sensitivity higher than previous observatories.
The main results of this survey and the relevant source catalogues are reported in 
several  papers (e.g. 
\citealp{bird04,bird06,bird07,bassani06,krivonos07,krivonos05,sazonov07,churazov07}).

The Burst Alert Telescope (BAT; \citealp{bat}) on board the
{\it Swift} observatory \citep{swift}, with its large field of view ($100^{\circ}\times60^{\circ}$ 
 half coded) and large detector area (a factor of 2 greater than ISGRI)
offers the opportunity for a large increase of the sample of objects that
contribute to the luminosity of the sky in the hard X-rays allowing for a 
substantial improvement of our knowledge of the AGN and of the cosmic 
hard X-ray background.
The first results on the BAT survey have been presented in \citet{markwardt05,ajello1,ajello3,tueller08}. 
The latter presents a catalogue of sources detected in the first 9 months of the
BAT survey data, identifying 154
extragalactic sources (129 at $|b|>15^{\circ}$).

In order to exploit the BAT survey archive, we developed the dedicated software
{\sc BatImager} \citep{segreto09}, independent from the one 
developed by the Swift-BAT team\footnote{http://heasarc.gsfc.nasa.gov/docs/swift/analysis/}.
In this paper we present the results obtained from the analysis
of 39 months of BAT sky survey.
The paper is organized as follows: in
Sect.~\ref{bat} we describe the BAT
telescope; in Sect.~\ref{datascreen} we describe the data
set and screening criteria; in Sect.~\ref{method} we present a brief
description of the code used for the analysis and illustrate our
analysis strategy. In Sect.~\ref{skycov_t} we describe the survey properties. 
The catalogue construction and the results are 
reported in Sect.~\ref{cat}. 
The last Section summarizes our results. The spectral 
properties of our extragalactic sample will be discussed in a forthcoming paper
(La Parola et al. 2009, in preparation).

The cosmology adopted in this work assumes $H_0=70$ km s$^{-1}$ Mpc$^{-1}$,
k=0, $\Omega_m=0.3$, and $\Lambda_0=0.7$. Quoted errors are at $1\sigma$
confidence level, unless otherwise specified.

%%%%%%%%%%%%%%%%%%%%%%%%%%%%%%%%%%%%%%%%%%%%%%%%%%%%%%%%%%%%%%%%%%%%%

\section{The BAT telescope\label{bat}}
The BAT, one of the three instruments on board the {\it Swift} observatory, 
is a coded aperture imaging camera consisting of a 5200 cm$^2$ array 
of  $4\times4$ mm$^2$ CdZnTe elements mounted
on a plane 1 meter behind a 2.7 m$^2$ coded aperture mask of 5 $\times$ 5 
mm$^2$ elements distributed with a pseudo-random pattern.
The telescope, operating in the 14--150 keV energy range  with a
large field of view (1.4 steradian half coded)  and a point spread function
(PSF) of 17 arcmin, is mainly devoted to the monitoring of a large fraction
of the sky for the occurrence of Gamma Ray Bursts (GRBs). The BAT provides their
position with the accuracy (1--4 arcmin) that is necessary to slew
the spacecraft towards a GRB position and bring the burst location inside the 
field of view of the narrow field instruments in a couple of minutes.
While waiting for new GRBs, it continuously collects spectral and
imaging information in survey mode, covering a
fraction between 50\% and 80\% of the sky every day. The data are immediately
made available to the scientific community through the public Swift data
archive\footnote{http://heasarc.gsfc.nasa.gov/cgi-bin/W3Browse/swift.pl}.

%%%%%%%%%%%%%%%%%%%%%%%%%%%%%%%%%%%%%%%%%%%%%%%%%%%%%%%%%%%%%%%%%%%%%

\section{Survey data set and screening criteria\label{datascreen}}
We have analysed the first 39 months of the BAT survey data archive,
from 2004 December to the end of 2008 February. 
The BAT survey data are in the form of Detector Plane Histograms 
(DPH). These are three dimensional arrays (two spatial dimensions, 
one spectral dimension) which collect count-rate data in 
(tipically) 5-minutes time bins for 80 energy channels.

The data were retrieved from the {\it Swift} public archive and screened out 
from bad quality files, excluding those files where the
spacescraft attitude was not stable (i.e., with a significant variation of
the pointing coordinates).
The resulting dataset was pre-analized (see Sect.~\ref{method}), in order to 
produce preliminary 
Detector Plane Images (DPI, obtained integrating the DPH along the spectral
dimension) from where the bright sources (S/N $> 8$) and background 
were subtracted; very noisy DPHs, i.e. with a standard deviation 
significantly larger than the average value where subtracted.
The list of bright sources detected in each DPH was used to identify and 
discard the files suffering from inaccurate
position reconstruction. After cross-correlating the position
of these sources with the ISGRI catalogue, the GRB positions and the newly 
discovered {\it Swift} sources documented in literature
\citep{markwardt05,ajello1,tueller08}, 
we discarded the files where:
\begin{itemize}
\item the bright sources in the BAT field of view are detected at
more than 10 arcmin from their counterpart position (due
to a star tracker loss of lock).
\item the reconstructed image of at least one bright source has a
strongly elongated shape (maybe due to an unrecognized
slew).
\end{itemize}

After the screening based on these criteria, the usable archive has a total
nominal exposure time of 72.7 Ms, corresponding to 91.2\% of the total survey
exposure time during the period under investigation.

%%%%%%%%%%%%%%%%%%%%%%%%%%%%%%%%%%%%%%%%%%%%%%%%%%%%%%%%%%%%%%%%%%%%%

\section{Methodology\label{method}}
In order to perform a systematic and efficient search for new hard 
X-ray sources, we  have developed the 
{\sc BatImager},  a dedicated software which produces an all-sky mosaic directly
from a list of BAT data files.
A complete and detailed description of the software and its 
performance is presented in \citet{segreto09}. Here we only report the details
of the procedure which are relevant to this work.

\subsection{The code}
The {\sc BatImager} integrates each single DPH
in a selected energy range, producing the corresponding DPI. 
A preliminary cleaning of the disabled and noisy pixels is performed, and the 
DPI is cross correlated with the mask pattern, in order to identify and 
subtract  bright sources (with S/N $>8$). Then the
background, modelled on a large scale from the analysis of the shadowgram residuals by performing a
Principal Component Analysis \citep{kendall80}, is subtracted. A further search for
 bad pixels is performed, obtaining the final map of all pixels to be
excluded in the following steps.
A further correction is applied to take into account differences in the detection
efficiency of single detector pixels, through a time/energy dependent 
efficiency map, built stacking all the processed DPI and equalizing the average
residual contribution for each pixel. 
 The original DPI, corrected for the efficiency map and cleaned for the 
bad pixels, is processed again, with all the contributions from the 
background and the bright sources identified in the previous steps
computed simultaneously, in order to correct for
cross-contamination effects. These contributions are subtracted from the
DPI, that  is then converted into a sky image, using the  Healpix 
projection \citep{HEALPix}. This projection provides an equal-area pixelization 
on a sphere and allows the generation of an all-sky map, 
avoiding the distortion introduced by other types of 
sky projections far from the projection center. This sky map is then corrected for
the occultation of Sun, Earth and Moon. The sky maps produced from each DPI are
added together, with the intensity in a given sky
direction computed from the contribution from all the sky images, each inversely
weighted for its variance in that direction. As described above, the bright 
sources and background were already subtracted from each single DPI; therefore
this all-sky mosaic contains only the residual sky contribution.
In order to correct for residual 
systematic effects (e.g. imperfect modelling of the source illumination pattern
or of the background distribution), the all-sky S/N map is  sampled on a 
scale significantly larger than the PSF: the local average S/N is
subtracted and its measured variance used to normalize the local S/N 
distribution. Finally, we obtain a
S/N map with zero average and unitary variance that can be used for a
blind source detection.

%%%%%%%%%%%%%%%%%%%%%%%%%%%%%%%%%%%%%%%%%%%%%%%%%%%%%%%%%%%%%%%%%%%%%

\begin{figure}
\centerline{\psfig{figure=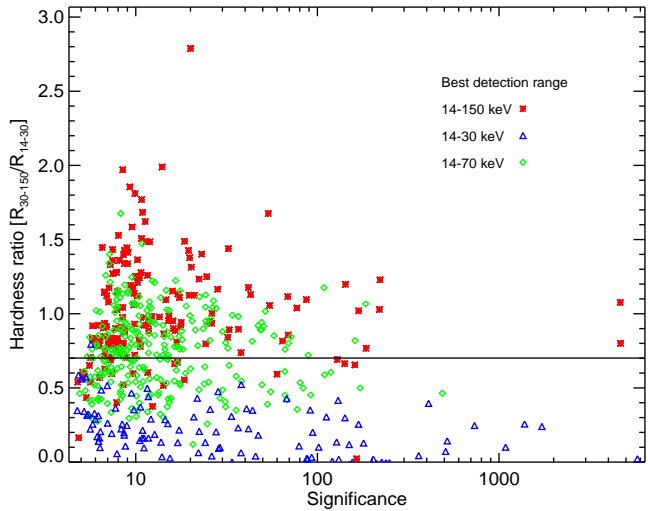,width=9cm}
            }
\caption{Hardness ratio [defined as R(30--150)/R(14--30)] of the sources 
detected with {\sc BatImager} as a function of the best detection significance.
Different symbols refer to the energy range where each source was detected at the
highest S/N. The solid line is the average hardness ratio value.
\label{hr}}

\end{figure}

\subsection{Detection strategy\label{detstrategy}}

We have created all-sky maps in three energy bands: 14--150 keV, 14--70 keV, 
14--30 keV. 
The source detection in the all-sky map is performed by searching
for local excesses in the significance map. 
The source position and its peak significance are then refined with
a fit restricted within a region of a few
pixels where the excess dominates over the noise distribution. 
Only detections with peak significance greater than 
4.8 sigma  are included in our list of detected sources. We found that this 
threshold represents the optimal value that maximizes the
number of detectable sources, maintaining at the same time an acceptable number
of spurious detections: taking into account the total number of pixels in the all 
sky map, the PSF and the Gaussian distribution of the noise, we expect 23
spurious
detections above our threshold in each energy band, due to statistical
fluctuations. Therefore, the total number 
of spurious detections will be between 23 and 69 (2.4\% to 7.2\% of the total number
of our detections, see below), the best case occurring if each
noise fluctuation above the threshold appears simultaneously in all three 
bands, the worst case occurring if each fluctuation appears only in one energy 
band. A few sources ($\sim 5\%$) detected with a significance slightly lower than our
threshold were included in the detection list because their S/N is significantly
larger than the negative excess (in modulus) of the local noise distribution.

The resulting detection catalogues (one for each of the three energy bands)
have been cross-correlated  and merged in a 
single catalogue: when  source candidates closer than 10 arcmin are present in the sky maps 
of different energy bands, they were reported in the merged catalogue as a single source
candidate. We obtain a final number 
of   962 source candidates (detected in at least
one of the three energy bands). We adopt as best source position
the one corresponding to the energy range with the highest detection
significance.

 We have evaluated the hardness ratio of the detected sources as 
Rate(30--150 keV)/Rate(14--30 keV) (the hard rate is evaluated as 
the difference between the count rates in the 14--150 and in the 14--30 energy bands).
In Figure~\ref{hr} we plot the hardness ratio as a function of the significance for each 
detected source, showing the energy range where the detection has the
highest significance. Repeating the detection
process in three energy bands optimizes the S/N for each source, and 
this yields better values for the source position, whose uncertainty 
scales inversely with the significance. Moreover, a significant subsample of
sources was detected in only one of the three energy bands ( 56 in the 14--150
keV energy band, 38 in the 14--30 keV energy band, 78 in the 14--70 keV energy 
band) demonstrating that
searching in different energy bands maximizes the number of detectable sources.

\begin{figure}
\centerline{\psfig{figure=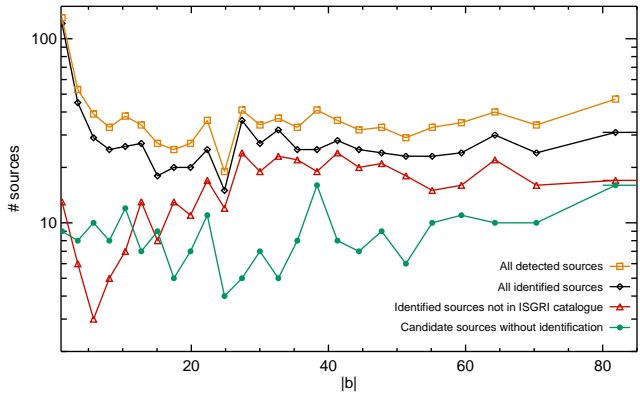,width=9cm}
            }
\caption{Distribution of the detected sources vs. Galactic latitude. Each
bin corresponds to a solid angle of $\sim0.50$ sr.
\label{profile}}

\end{figure}

Figure~\ref{profile} shows that the distribution of the detected sources
(orange squares) vs. Galactic latitude flattens for $|b| > 5$, 
which we shall hereon consider our operational definition of the Galactic plane.
%%%%%%%%%%%%%%%%%%%%%%%%%%%%%%%%%%%%%%%%%%%%%%%%%%%%%%%%%%%%%%%%%%%%%

\subsection{Identification strategy\label{id}}

The identification of the counterpart of the BAT detections was performed 
following three different strategies.
\\

{\it A}. The position of each of the 962 detected excesses  was cross-checked  
with the 
coordinates of the sources included in the INTEGRAL General Reference 
Catalogue\footnote{http://isdc.unige.ch/?Data+catalogs} (v. 27), 
 that contains 1652 X-ray emitters,
and with the coordinates of the counterpart of the 48 new identifications of  
BAT sources already 
published \citep{markwardt05,tueller08,ajello1,ajello3} and not 
included in the above catalogue. We adopted as counterpart a source within
a radius $R=8.4$ arcmin from the BAT position (4 standard deviations error circle
for a source detection at 4.8 standard deviations,
\citealp{segreto09}). With this method we obtain 458 identifications, 
295   with $|b|>5^{\circ}$. The choice of the error radius is 
strategic to maximize the associations and
keep the number of spurious associations to a negligible level. 
The number of spurious
identifications due to chance spatial coincidence has been evaluated 
using the following expression:

\begin{equation}\label{fake1}
N_{\rm sp}=\frac{N\times A_R}{A}\times{N_{\rm cat}}
\end{equation}

where, A$_R$ is the selected error circle area, A is the total
sky area under investigation, $N$ and $N_{\rm cat}$ are the number of BAT 
detections and of candidate counterparts in A. The above formula
assumes both source distributions to be uniform over the sky. In order to take 
into account the higher density of sources on the Galactic plane we have
divided the sky into two regions: $|b|\leq 5^{\circ}$ (the Galactic plane,
with $N= 190$, $N_{\rm cat}=651$) and $|b|>5^{\circ}$ 
($N=772$, $N_{\rm cat}=1049$).  The number of expected spurious identifications is 
2.1 within 
$|b|<5^{\circ}$ and 1.3 elsewhere. As the assumption of uniform distribution 
could be only a crude approximation, we have verified the evaluation of expected
spurious associations with an  alternative method: we produced a set of 962
coordinate pairs by inverting the position of the detected excesses with
respect to the Galactic reference system and cross-correlated these positions
with the  INTEGRAL General Reference Catalogue extended with the published BAT
identifications. We obtained 3 spurious associations, in full agreement with 
the value obtained in Eq.\ref{fake1}.
\\

{\it B}. We have searched for observations from {\it Swift}/XRT containing  
the remaining (504) unidentified excesses in their field. We found {\it Swift}/XRT 
observations for 186 BAT source candidates.
Source detection inside these X-ray images was 
performed using {\sc XIMAGE} (v4.4).
When a source was detected inside a $6.3$ arcmin error circle (99.7\%  
confidence level for a source detection at 4.8 standard deviations,
\citealp{segreto09}) we first 
checked for its  hardness ratio in the 0.3--10 keV
range (with  3 keV as a common boundary of the two ratio bands) and for its 
count rate above 3 keV.
We identified a source as the counterpart of a BAT detection if at 
least one of the above conditions was
satisfied: hardness ratio $>0.5$, count rate above 3 keV 
$>5\times10^{-3}$ c s$^{-1}$. 
In seven cases where two candidates, satisfying at least one of the threshold 
conditions, were found inside the BAT error circle, we chose as counterpart 
the closest source to the BAT position.
With this method we identified 170 source counterparts.
 In order to evaluate the
number of expected spurious identifications we collected a large sample of XRT
observations of GRB fields, using only late follow-ups (where the GRB afterglow
has faded) with the same exposure time
distribution as the XRT pointings of the BAT sources. We searched for
sources within a 6.3 arcmin error circle centered at the nominal pointing
position in each of these fields, excluding any GRB residual afterglow, and
satisfying at least one of the above threshold conditions. We detected
7 sources, therefore, the number of expected spurious identifications is 
consistent with the number of multiple XRT detections inside the BAT error
circle. 
We also searched for field observations with other X-ray instruments
({\it XMM-Newton}, {\it Chandra}, {\it BeppoSAX}), finding 25 identifications, out of 30 
pointings. Given the low number of available fields, the number of expected
spurious identifications within this sample is irrelevant.  
\\

{\it C}. For the remaining unidentified sky map excesses (309) we searched for spatial 
coincidence inside an error circle of 4.2 arcmin radius ($90\%$ confidence 
level for a source detection at 4.8 standard deviations, \citealp{segreto09}) 
with sources included in the SIMBAD catalogues. The size of the search
radius was fixed to 4.2 arcmin in order to have a negligible number of spurious
identifications (see below). We restricted our search to
the following SIMBAD object classes: Cataclysmic variable (CV), High mass 
X-ray binaries (HXB), Low mass X-ray binaries (LXB), Seyfert 1 (Sy1), Seyfert 
2 (Sy2), Blazar and BL~Lac (Bla, BLL),
LINERs (LIN), for a total of 22425 objects in the SIMBAD database. 
This strategy allowed us to identify  
92 detections, with only one source at low Galactic latitude ($|b|<5^{\circ}$). 
The number of expected spurious identifications was  evaluated with the two 
methods described for the strategy {\it A}. According to Eq.\ref{fake1} we 
expect 0.03 spurious identifications 
within $|b|<5^{\circ}$ (20 BAT detections and 391 Simbad sources in the 
classes of interest) and 2.7 elsewhere (289 BAT
detections and 22034 Simbad sources); using the set
of 309 coordinate pairs obtained inverting with respect to the Galactic center 
the positions of the sources in our sample we find 3 spurious associations,
consistent with the first method.
The cross correlation between unidentified sky excess and the SIMBAD catalogue of QSOs 
was treated separately because the coincidence error circle of 4.2 arcmin radius results
in a high number of spurious associations (9 out of 17 associations). A radius of 2 arcmin
allowed us to identify  9 sources as QSOs, and to optimize the ratio between the total number of associations and the expected number of
spurious associations ($\sim2$).\\

\begin{figure}
\centerline{\psfig{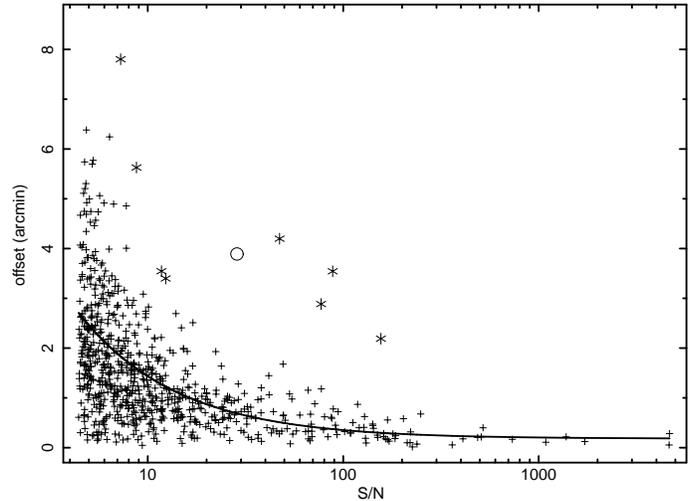} }
\caption{ Offset between the BAT position and the counterpart position as a function 
of the detection significance. A few values are far from the overall distribution:
those marked with a star (sources number 535, 564, 565, 570, 571, 574, 584 and 586 in
Table~\ref{srctab}) are in crowded field and the reconstructed sky position 
suffers from the contamination of the PSF of the nearest sources; the one marked with a circle
is an extended source (Coma Cluster). The solid line represents the fit to the data
(excluding the few outliers) with a power
law. \label{offset}}
\end{figure}

In Fig. \ref{offset} we report the offsets of each BAT source with respect to its identified 
counterpart as a function of the detection significance (S/N). 
The offset vs. the detection significance can be
modeled with a power-law plus a constant. The best fit equation we obtained is the
following:

\begin{equation}
\textup{ Offset}{\rm (arcmin)} = (9.1\pm 1.6) \times \textup{(S/N)}^{-0.93\pm 0.09} + (0.21\pm 0.03)
%\ \ \ \ {\rm [arcmin]}
\label{eq:off_survey}
\end{equation}

The  constant in Eq. \ref{eq:off_survey}  represents the sistematic due to a residual 
boresight misalignment. At the detection threshold of 4.8 standard deviations the average
offset is $\sim 2.6$ arcmin.

\begin{figure}
\centerline{\psfig{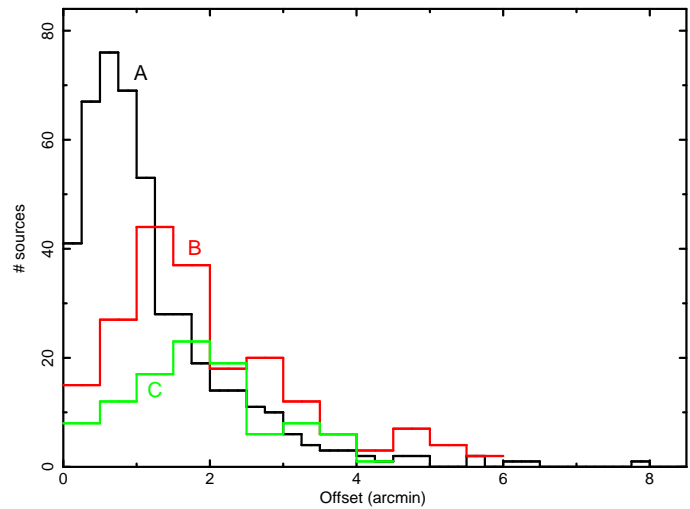} }
\caption{Distribution of the identified sources for each identification strategy
(Sect.~\ref{id}) as a function of the offset between the BAT position and the counterpart
position. 
  \label{histoid}}
\end{figure}

Fig. \ref{histoid} shows the distribution of the identified sources for each identification
strategy as a function of the offset between the BAT position 
and the counterpart position. The peak of the distribution is at lower offset for strategy
{\it A} because the sample of the sources identified with this strategy contains the
brightest objects. The  peak of the distribution relevant to strategy {\it B} is at a lower
offset with respect to the distribution of strategy  {\it C} because the XRT follow-up
observations were performed on the more significant still unidentified source candidates.

All the identifications obtained with the three strategies (754) were merged in 
the
final catalogue reported in Table~\ref{srctab} (see Section~\ref{cat}) where a 
flag indicates the identification method for each source. Figure~\ref{profile}
shows the distribution of all identified sources (black diamonds) as a function
of the Galactic latitude.

A set of 208 detections could not be associated  with a counterpart. 
These source candidates  have detection significance between 4.8 and
14 standard deviations and flux in the 14--150 keV band between $6.7\times
10^{-12}$ and $2.7\times10^{-11}$ \ferg. 33 sources out of 208 are detected
in all the three enegy bands and 63 in two energy bands. The unidentified 
detections are distributed quite uniformly in the sky (Figure~\ref{profile}, 
green circles), with 190 sources out of
208 located above the Galactic plane ($|b|>5^{\circ}$).

%%%%%%%%%%%%%%%%%%%%%%%%%%%%%%%%%%%%%%%%%%%%%%%%%%%%%%%%%%%%%%%%%%%%%

\begin{figure}
\centerline{\psfig{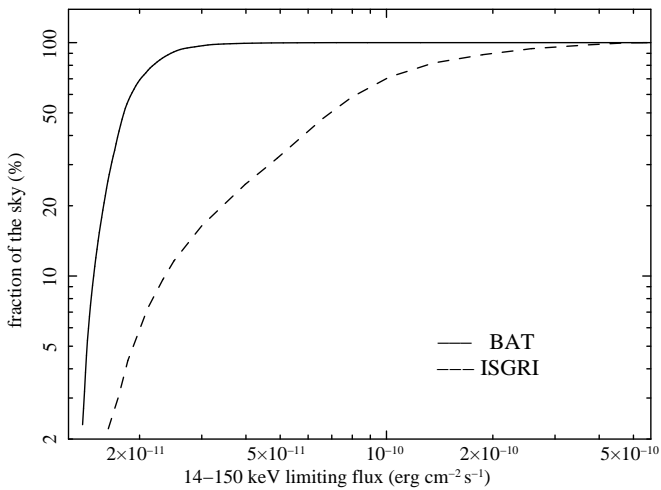}
            }

\caption{Fraction of the sky covered by the {\it Swift}-BAT and INTEGRAL-ISGRI 
surveys vs. limiting flux. \label{skycov}}
\end{figure}

\begin{figure*}
\centerline{\psfig{figure=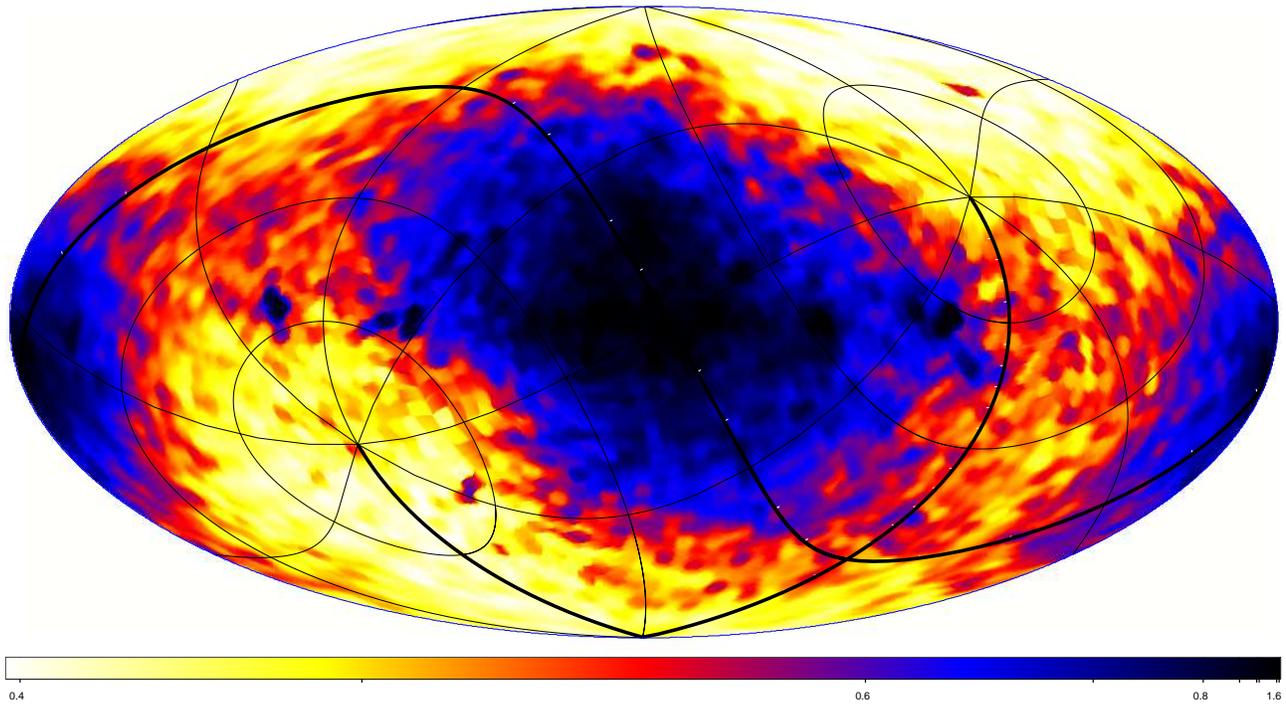,width=17cm}
            }

\caption{Map of the limiting flux (in mCrab) of the 39-months BAT-survey data in the
14--150 keV band, projected in Galactic coordinates, with the
ecliptic coordinates grid superimposed (the thick lines represents the ecliptica
axes). The scale on the colorbar is in mCrab.
\label{flim}}
\end{figure*}

\section{Sky coverage and limiting flux\label{skycov_t}}
Figure~\ref{skycov} shows the sky coverage, defined as the fraction of the sky 
covered by the 
survey as a function of the detection limiting flux. The limiting flux for a given 
sky direction is calculated  by multiplying the local image noise by a 
fixed detection threshold of $5$ standard deviations. This threshold, higher 
than the one adopted for source detection (Sect.~\ref{detstrategy}), was 
used to compare the 
BAT sky coverage with those produced with the INTEGRAL data survey. 
The large BAT field of view, the  large 
geometrical area together with the Swift pointing distribution, covering the
sky randomly and uniformly according to the appearance of GRBs,
has allowed the achievement of an unprecedented 
sensitive and quite uniform sky coverage.
The 39 months BAT survey covers 90\% of the sky  down to a
flux limit of  $ 2.5\times10^{-11}$ 
\ferg (1.1 mCrab), and  50\% of the sky down to $1.8\times10^{-11}$ \ferg 
(0.8 mCrab). In the same figure the BAT sky coverage 
is compared with that of  INTEGRAL/ISGRI after 44 months of observation 
\citep{krivonos07}. 

Figure~\ref{flim} shows the limiting flux map in galactic Aitoff 
projection, with the ecliptic coordinates grid superimposed. 
The minimum detection limiting flux is not fully uniform on the sky: 
the Galactic center  and the ecliptic plane are characterized by a 
worse sensitivity due to high contamination from  intense 
Galactic sources and to the observing constraints of the {\it Swift} spacecraft. 
The highest flux sensitivity is achieved near the ecliptic poles 
where a detection flux limit of about $1.1\times10^{-11}$\ferg~ is reached 
($\sim0.5$ mCrab).

%%%%%%%%%%%%%%%%%%%%%%%%%%%%%%%%%%%%%%%%%%%%%%%%%%%%%%%%%%%%%%%%%%%%%

\section{The 39-months catalogue\label{cat}}
The complete catalogue of the sources identified in the first
39 months of BAT survey data is reported in Table~\ref{srctab}.
The table contains the following information:
\begin{itemize}
\item Palermo BAT Catalogue (PBC) name of the source (column 2), built from the BAT coordinates 
with the precision of 0.1 arcmin on RA.
\item Counterpart identification (column 3) and source type (column 4) 
coded according to the nomenclature used in SIMBAD.
\item RA and Dec of the BAT source in decimal degrees (columns 5, 6).
\item Error radius (column 7), offset with respect to the counterpart position 
(column 8) and significance (column 9), as obtained in the energy band with the 
highest significance (a flag in column 14 indicates the energy range with the 
maximum significance).
\item Flux in the widest band of detection, averaged over the entire
survey period (column 10). For most of the sources 
this is 14--150 keV. In the other cases a flag in column 14 indicates the 
appropriate band.
In order to convert count rates into fluxes we derived a conversion factor for
each of the three bands using the corresponding Crab count rate and the Crab
spectrum used for BAT calibration purposes, as reported in the BAT calibration status
report\footnote{http://swift.gsfc.nasa.gov/docs/swift/analysis/bat\_digest.html\#calstatus}.
\item Hardness ratio defined as Rate[30--150 keV]/Rate[14--30 keV], 
where the hard rate is evaluated as the difference between the count rates in
the 14--150 and in the 14--30 energy bands (column 11).
\item Redshift of the extragalactic sources (column 12), from the SIMBAD
database (or NED, for the few cases that were not reported in SIMBAD).
\item Log of the rest frame luminosity in the 14--150 keV band for extragalactic objects 
(column 13), calculated using the
luminosity distance for sources with redshift $>0.01$, and using the distance
reported in the Nearby Galaxies Catalogue (NBG, \citealp{tully88}) or NED, for 
the 
few cases that were not reported in the NBG catalogue, for sources with redshift
$< 0.01$.
\item Flag column (column 14) with information on: energy band with the highest
significance (A), energy band used for the calculation of the flux (B), flag 
for already known hard
X-ray sources (C), position with respect to the Galactic plane ($|b|<5^{\circ}$,
D), strategy used
for the identification (E, see Sect.~\ref{id})
\end{itemize}

\subsection{Statistical properties of the catalogue}

\begin{figure*}
\centerline{\psfig{figure=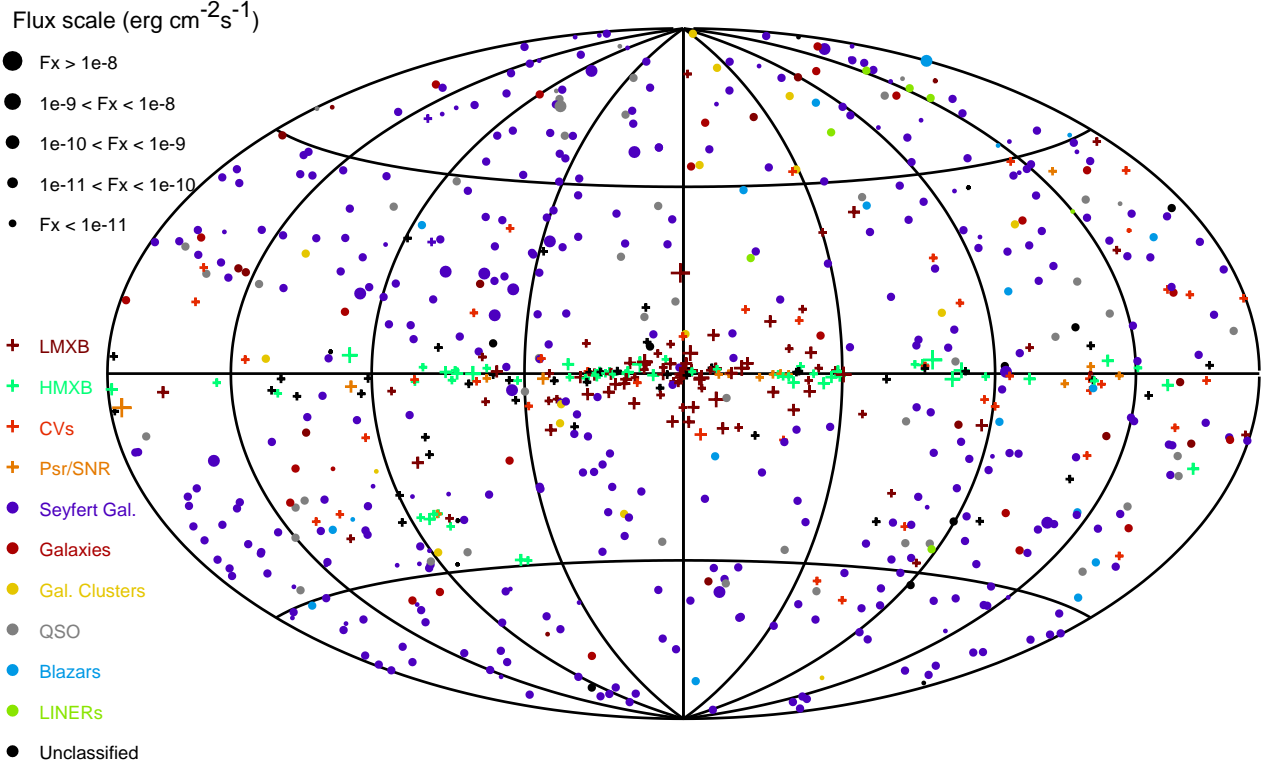,width=18cm}
           }
 
\caption[]{Map of the sources we detect in the BAT survey data (Galactic coordinates). Different colors
denote different object classes, as detailed in the legend. The size of the
symbol is proportional to the source flux in the 14-150 keV band.
\label{aitoff}}
\end{figure*}

Table~\ref{types} details the distribution of the 754 sources in our
catalogue among different object classes: $\sim 69$\% of the catalogue
is composed of extragalactic objects, $\sim 27$\% are Galactic 
objects, $\sim 4$\% are already known X-ray or gamma ray emitters whose
nature is still to be determined. Figure~\ref{aitoff} shows the distribution of all
the sources in our catalogue, colour-coded according to the object class,
with the size of the symbol proportional to the 14--150 keV flux (for those
sources not detected in the 14--150 keV band the flux in the widest band of
detection has been extrapolated to the 14--150 keV range using the BAT Crab 
spectrum). 

\begin{table}
\begin{tabular}{l r r}
\hline\hline
Class                & \# of sources& \% in the Catalog \\ \hline
LXB                  & 76          & 10.1\%\\
HXB                  & 64          &  8.5\% \\
Pulsars              & 10          &  1.3\% \\
SN/SNR               & 5           &  0.7\% \\
Cataclysmic variables& 46          &  6.1\% \\
Stars                & 5           &  0.7\% \\ 
Molecolar Cloud      & 1           &  0.1\%   \\ \hline 
Galactic (total)     & 207         &  27.5\%\\  \hline
Seyfert 1 galaxies   & 235         & 31.2\% \\ 
Seyfert 2 galaxies   & 131         & 17.4\% \\ 
LINERs               & 7           & 0.9\%  \\
QSO                  & 14          & 1.8\%  \\
Blazars              & 71          & 9.4\%  \\
Galaxy clusters      & 18          & 2.4\%  \\
Normal galaxies      & 27          & 3.6\%  \\
Unclassified AGN     & 16          & 2.1\%  \\ \hline
Extragalactic (total)& 519         & 68.8\% \\ \hline
Other types          & 28          & 3.7\%\\ \hline
\end{tabular}
\caption{Classification of the known sources detected in the BAT survey. 
{\it Other types}
includes all sources that have a catalogued counterpart but have not
been classified yet.\label{types}}
\end{table}

\begin{figure}
\centerline{\psfig{figure=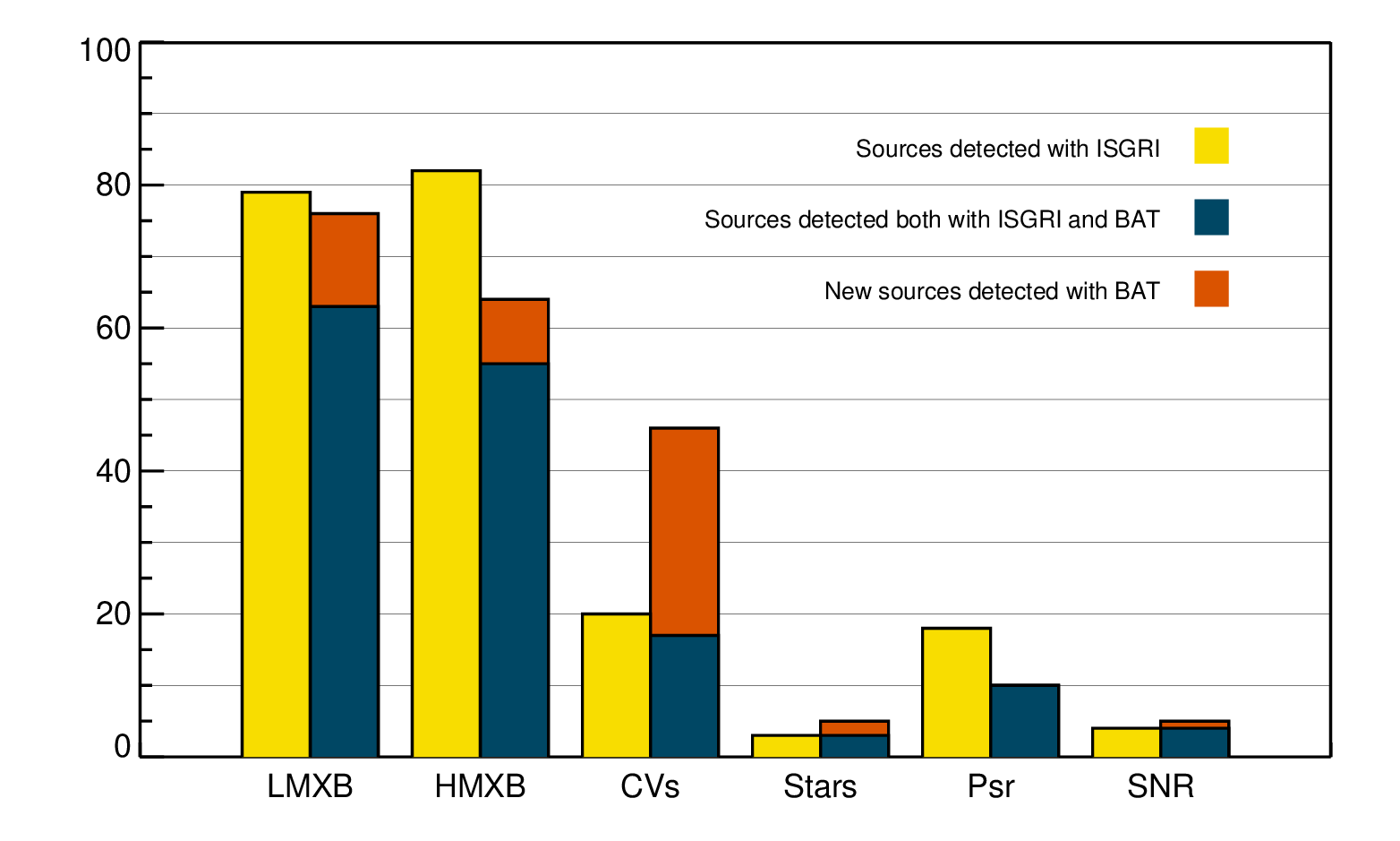,width=9cm,angle=0}
            }
\centerline{\psfig{figure=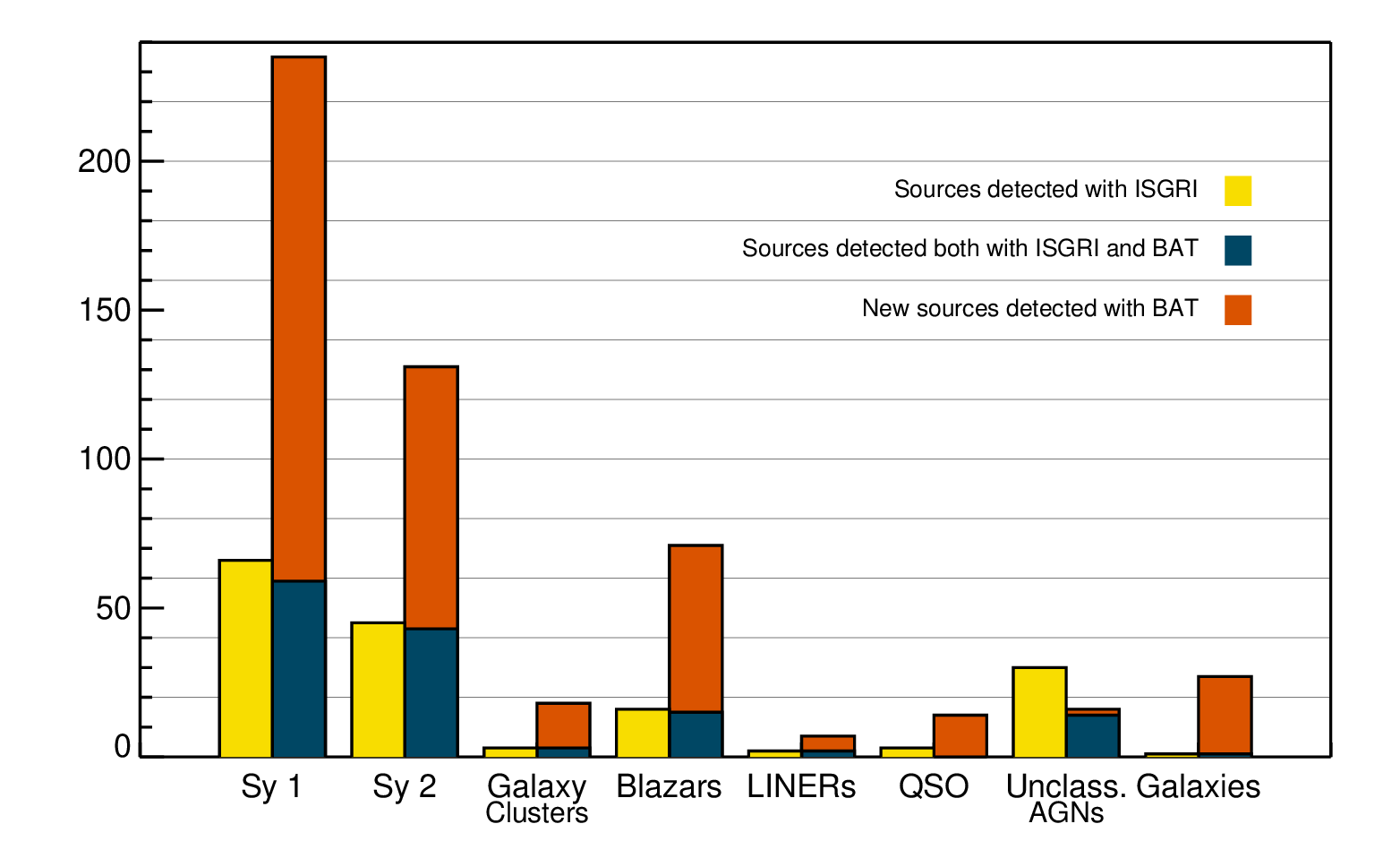,width=9cm,angle=0}
            }

\caption[]{Comparison between the sources in our catalogue and those 
reported in the third ISGRI catalogue \citep{bird07}. Top: Galactic sources.
Bottom: Extragalactic sources.
\label{histo}}
\end{figure}

We have compared this distribution with the third ISGRI catalogue
\citep{bird07}. The results are plotted in Figure~\ref{histo}. We find 
a dramatic improvement in the detection of extragalactic objects, both in
the nearby Universe (normal galaxies, LINERs) and at higher distances (Seyfert
galaxies, QSO, clusters of galaxies). As expected from the sky coverage achieved by the BAT
survey data (Figure~\ref{skycov}), most of our identified sources have a
flux below $1\times10^{-10}$  erg s$^{-1}$ cm$^{-2}$ and are located outside the
Galactic plane. We also detect many Galactic sources 
that are not included in the ISGRI catalogue, most of which are cataclysmic variables
 and  X-ray binaries. 
This can be explained  in part with the different pointing strategy of the two
instruments. Hovewer, Figure~\ref{profile} shows that, although most
of our newly identified sources (red triangles) are above the Galactic plane, where
the ISGRI exposure is low, we also detect
a few sources on the Galactic plane most
of which we identify as X-ray binaries (1E 1743.1--2852, GRO 1750--27, 
SAX J1810.8--2609 and
XTE J1856+053). We have verified that their detections are due to a
transient intense emission observed in the large FoV of BAT.

We detect emission from 18 clusters of galaxies. 
We verified that for 17 of them the spectral 
distribution in the 14--150 keV band is consistent with
the tail of a thermal emission with kT $\sim10$ keV without evidence for
the presence of a hard non-thermal emission. Only for Abell~2142 we find evidence
for a power law component that could be ascribed to the AGN content of the cluster.

%%%%%%%%%%Sono arrivata qua%%%%%%%%
%%%%%%%%%%%%%%%%%%%%%%%%%%%%%%%%%%%%%%%%%%%%%%%%%%%%%%%%%%%%%%%%%%%%
\subsection{The extragalactic subsample\label{extragal}}
\begin{figure}
\centerline{\psfig{figure=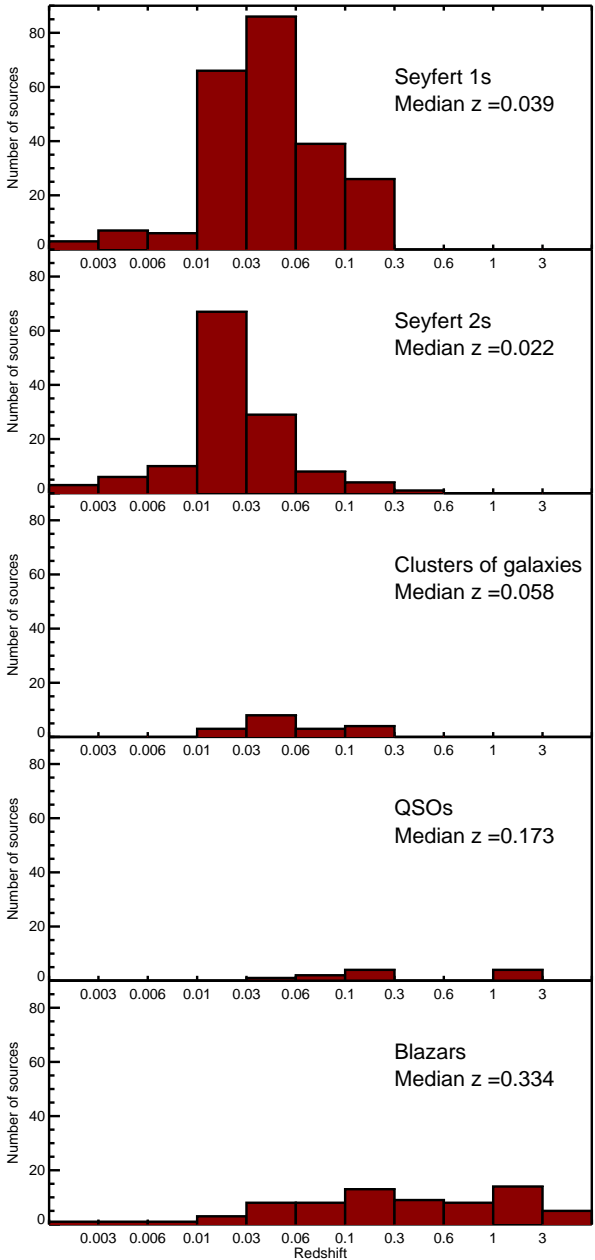,width=9cm}
            }
%\centerline{\psfig{figure=lumz.ps,width=9cm}
%            }

%\caption[]{Top: Redshift distribution of the extragalactic sources in the BAT survey
%catalogue. Bottom: Luminosity in the 14-150 keV band vs redshift for different
%class of extragalactic sources
%}
\caption[]{Redshift distribution of the extragalactic sources in the BAT survey
catalogue for different classes of extragalactic sources.
\label{z_dist}}
\end{figure}

The catalogue contains 519 extragalactic objects. Figure~\ref{z_dist} shows 
the distribution of the redshift within our sample for the main 
classes of extragalactic objects. Most of the emission-line AGNs are located 
at $z<0.1$, but we also 
detected a few Seyfert 1 galaxies at larger redshift (up to $\sim 0.29$). Seyfert 2
galaxies are detected up to $z\sim 0.4$. Blazars are detected up to $z\sim 3.7$,
and QSOs are detected up to $z\sim 2.4$.

%\subsection{Completeness of the sample\label{vvmaxtest}}

\begin{figure}
\centerline{\psfig{figure=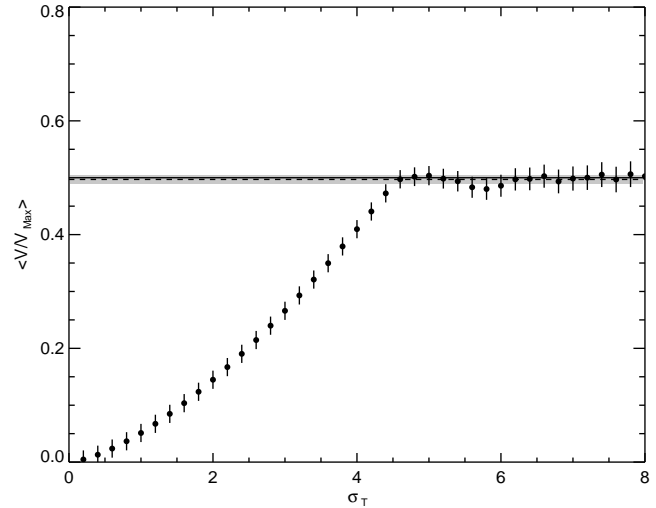,width=9cm}
            }

\caption[]{$<V/V_{\rm Max}>$ vs significance for our sample of extragalactic
sources. The solid line is the expected value (0.5), the dashed line is the
average value for $S/N>4.5\sigma$, the shaded area covers the $1\sigma$ error
for the average value.
\label{vvmax}}
\end{figure}

We verified the completeness of our sample of  366 emission line galaxies 
(i.e. the significance limit down to
which we are including in the sample all objects above a given flux limit) using
the $V/V_{\rm Max}$ test \citep{schmidt68,huchra73}. This method was developed
to test the evolution of complete samples of objects, but can be also used to
test the completeness of non-evolving samples.
For each source, $V$ is the volume enclosed by the object distance, while
$V_{\rm Max}$ is the volume corresponding to the maximum distance where the object
could be still revealed in the survey (and thus depends on the limiting flux in
the direction of the object). In case of no evolution the expected value of
$<V/V_{\rm Max}>$, averaged over the entire sample, is 0.5. 
We assume the hypothesis of no evolution and uniform distribution in the local 
Universe. For each source in the sample, and for each significance level tested for
completeness ($\sigma_{\rm T}$), we compute the quantity $V/V_{\rm Max}$ as
$[F/\sigma_{\rm T}\Delta F)]^{-3/2}$, where $F$ is the flux of the source and $\Delta F$
its 1 standard deviation uncertainty. $<V/V_{\rm Max}>$ is obtained averaging $V/V_{\rm Max}$ over
the number N of all sources in the sample detected with a significance higher 
than $\sigma_{\rm T}$, and its error is $1/12N$. Figure~\ref{vvmax} shows the results
of this test: the distribution becomes constant at $S/N\gtrsim 4.5\sigma$, with a mean
value of $0.497 \pm0.007$, consistent with the expected value of 0.5. Thus we
can confidently assume that our sample  is complete down to our adopted
significance threshold of $4.8\sigma$.

%%%%%%%%%%%%%%%%%%%%%%%%%%%%%%%%%%%%%%%%%%%%%%%%%%%%%%%%%%%%%%%%%%%%%

\subsection{$\log(N)-\log(S)$ distribution\label{lognlogs_t}}

The $\log(N)-\log(S)$ distribution was evaluated
by summing the contribution of all the detected sources firmly identified 
with extragalactic objects (Table~\ref{srctab}) and all the unidentified
detections. We selected only sources with $|b|>5^{\circ}$: Figure~\ref{profile} 
(orange squares) shows that the detection distribution is uniform above this
Galactic latitude limit.
The cumulative distribution is weighted by the area in which these sources 
could have been detected.
The following formula has been applied:

$$
N(>S)=\sum_{S_i>S} \frac{1}{\Omega_i},
$$

\noindent where $N$ is the total number of detected sources
with fluxes greater than $S$,
$S_i$ is the flux of the $i$-th source and $\Omega_i$ is the
sky coverage associated to the flux $S_i$ (Figure~\ref{skycov}).

In order to avoid the presence of systematic errors in the
determination of the $\log(N)-\log(S)$
arising because of spurious source detections and to the large relative 
uncertainty on the sky coverage at the lower end of the flux scale, 
we limited the construction of the $\log(N)-\log(S)$ to fluxes greater than 
$\sim1.5\times10^{-11}$ \ferg.
The resulting $\log(N)-\log(S)$ distribution contains
 330 sources ( 14 unidentified) and covers a flux range
up to $3\times10^{-10}$ erg s$^{-1}$ cm$^{-2}$.

We applied a linear least-square fit to derive the slope
of the $\log(N)-\log(S)$ distribution assuming 
a power law in the form $N(>S)=K \times (S/S_0)^{-\alpha}$, where $S_0$ is set
to $1\times10^{-11}$ \ferg.
The fit gives a value of $\alpha=1.56 \pm 0.06$ and a normalization of
$570\pm24$ sources with flux greater than $10^{-11}$ \ferg, corresponding to a
density of $(1.38\pm 0.06)\times 10^{-2}$ deg $^{-2}$.
The single power-law model is found to give
an acceptable description of the data ($\chi^2 = 0.65$; 31 dof) with a 
slope consistent with an Euclidean distribution.

The presence of spurious detections in the sample of
unidentified sources could introduce a systematic effect both in the
slope and in the normalization of the $\log(N)-\log(S)$. We expect
between 23 and 69 spurious detections due to statistical
fluctuations (see Sect.4.2), that correspond to a percentage between
$\sim 11$ and $\sim 33$ \% in the sample of the $\sim 208$ unidentified
sources. This means that  2-5 unidentified sources among those
used in the fit of the $\log(N)-\log(S)$ could be spurious.We have
checked that their contribution does not introduce any significant
systematics in the best fit values.

The integrated flux is $\sim 4.5\times10^{-13}$
erg cm$^{-2}$ s$^{-1}$ deg$^{-2}$ corresponding to $\sim$1.4\% of the 
intensity of the
X-ray background in the 14--170 keV energy band as measured by HEAO-1
\citep{gruber99}.\\

We have compared this $\log(N)-\log(S)$ law with the one derived from the
{\it INTEGRAL}  data \citep{krivonos07} in the $17-60$ keV band.
To convert our $\log(N)-\log(S)$ into the $17-60$ keV band we use the Crab spectral 
parameters derived by the INTEGRAL  
analysis \citep{laurent03}. 
We find a slope of $\alpha=1.62 \pm 0.08$ and a normalization of 
$240\pm 12$ sources with flux higher than 1 mCrab, corresponding to a density of 
$(5.8\pm 0.3)\times 10^{-3}$ deg$^{-2}$.
These parameters are in full agreement with those reported by
\citet{krivonos07}.
\begin{figure}
\centerline{\psfig{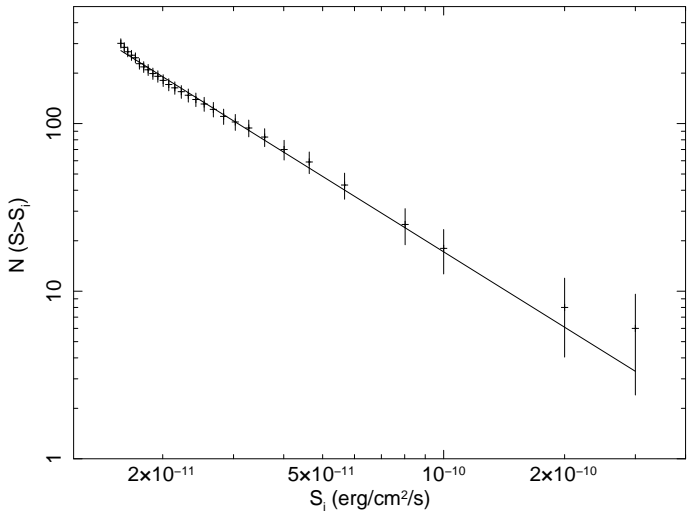}
            }

\caption[]{log(N)-log(S) distribution for the BAT extragalactic sources.
\label{lognlogs}}
\end{figure}

%%%%%%%%%%%%%%%%%%%%%%%%%%%%%%%%%%%%%%%%%%%%%%%%%%%%%%%%%%%%%%%%%%%%
\section{Conclusions\label{concl}}
We have analyzed the BAT hard X-ray survey data of the first 39 months of the
{\it Swift} mission. To this purpose we developed a dedicated software
\citep{segreto09} that performs data reduction, background subtraction, mosaicking 
and source detection on the BAT survey data. This software is completely
independent from the one developed by the {\it Swift}-BAT team. It is a single 
tool that provides all
the products relevant to the BAT survey sources (e.g. images, spectra, 
lightcurves). 

The large BAT field of view, the  large 
geometrical area, and the {\it Swift} pointing strategy
have allowed to obtain an unprecedented, 
very sensitive and quite uniform sky coverage
that has provided a significant increase of sources
detected in the hard X-ray sky. The survey flux limit is $2.5\times10^{-11}$ 
\ferg (1.1 mCrab) for 90\% of the sky and $  1.8\times10^{-11}$ \ferg (0.8 mCrab)
for 50\% of the sky.

We have derived a catalogue of  754 identified sources detected above a 
significance threshold of 4.8 standard deviations. The association of these
sources with their counterparts has been performed in three alternative
strategies: cross-correlation with the INTEGRAL General Reference 
Catalogue and with previously published BAT catalogues 
\citep{markwardt05,tueller08,ajello1}; analysis of soft X-ray field observations
with {\it Swift}-XRT, {\it XMM-Newton}, {\it Chandra}, {\it BeppoSAX}; cross-correlation with
the SIMBAD catalogues of Seyfert Galaxies, QSOs, LINERs, Blazars, Cataclysmic Variables,
X-ray binaries. The expected total number of spurious identifications is 
negligible. 
A set of  208 detections are not associated  with a counterpart, yet. These
candidate sources will be object of a follow-up campaign with {\it Swift}-XRT 
in the immediate future.  

The extragalactic sources represents $\sim 69$\% of our catalogue 
( 519 objects), $\sim 27$\% are Galactic objects, $\sim 4$\% are already known X-ray 
or gamma ray emitters whose nature is still to be determined. Compared with the 3rd
ISGRI catalogue \citep{bird07}, we identify  176 more Seyfert galaxies,  26 more normal
galaxies, 13 more galaxy clusters,  13 more QSO,  57 more Blazars and 5 more LINERs. 
The redshift limit for the detected emission line AGNs is $\sim 0.4$, with  31
objects with 
$z>0.1$.  Blazars and QSOs are detected up to $z\sim 3.7$ and $z\sim 2.4$,
respectively.
Among the Galactic sources we significantly increase the number of cataclysmic 
variables detected in the hard X-ray band ( 29 new objects). We also detect 22 X-ray 
binaries that are not
included in the ISGRI catalogue, even though the total number of X-ray binaries we
detect is lower than the sample included in the ISGRI catalogue.

Based on the extragalactic sources sample and on the achieved sky coverage, we 
have evaluated the $\log(N)-\log(S)$ distribution for fluxes higher than 
$1.5\times 10^{-11}$ \ferg. The slope $1.55\pm0.06$ is consistent with an 
Euclidean distribution.  We estimate that 
the total number of extragalactic sources at $|b| > 5^{\circ}$ and flux greater 
than $1.0\times 10^{-11}$ \ferg~ is $\sim566$. Converting this $\log(N)-\log(S)$
into the 17--60 keV band, our results are in full agreement with those reported
by \citet{krivonos07} for the INTEGRAL survey.
The integrated flux of this extragalactic sample is $\sim1.4\%$ of the Cosmic
X-ray background in the 14--150 keV range \citep{gruber99,churazov07,frontera07,
ajello-aph}.

Forthcoming papers will be focussed on the detection of transient sources,
spectral properties of the extragalactic sample, updates of the catalogue.

%%%%%%%%%%%%%%%%%%%%%%%%%%%%%%%%%%%%%%%%%%%%%%%%%%%%%%%%%%%%%%%%%%%%%%%%
\begin{acknowledgements}

G. C. acknowledges B. Sacco and M. Ajello for useful discussions that helped
to inprove this paper.
This research has made use of NASA's Astrophysics Data System Bibliographic Services,  
of the SIMBAD database, operated at CDS, Strasbourg, France, as well as of the 
NASA/IPAC Extragalactic Database (NED), which is operated 
by the Jet Propulsion Laboratory, California Institute of Technology, under contract with 
the National Aeronautics and Space Administration. 
%This work was supported by MIUR grant 2005-025417 and contract ASI/INAF
This work was supported by contract ASI/INAF
I/011/07/0.

\end{acknowledgements}

%%%%%%%%%%%%%%%%%%%%%%%%%%%%%%%%%%%%%%%%%%%%%%%%%%%%%%%%%%%%%%%%%%%%%%%%% BIBLIO 
\bibliographystyle{aa}
{}

%\bibliography{survey}

\clearpage
\scriptsize
\onecolumn
\begin{landscape}
\begin{longtable}{rlllrrrrrlrrrl}
\caption{BAT survey new sources: identification\label{srctab}}\\
\hline\hline
   & PBC name & ID & Type$^{\star}$ & RA  & Dec & Error radius& Offset & SNR & Flux$^{\dag}$& Hardness ratio      &Redshift&$\log{L_{14-150}}$  &Flag$^{\ddag}$\\ 
   &            &    &      &(deg)&(deg)&(arcmin)  &(arcmin)&     &(\ferg)&$(R_{30-150}/R_{14-30})$&        &(erg s$^{-1})$ &A B C D E   \\ 
\hline
\endfirsthead
\caption{continued.}\\
\hline\hline
   & PBC name & ID & Type$^{\star}$ & RA  & Dec & Error radius& Offset & SNR & Flux$^{\dag}$& Hardness ratio      &Redshift&$\log{L_{14-150}}$  &Flag$^{\ddag}$\\ 
   &            &    &      &(deg)&(deg)&(arcmin)  &(arcmin)&     &(\ferg)&$(R_{30-150}/R_{14-30})$&        &(erg s$^{-1})$ &A B C D E   \\ 
\hline
\endhead
\hline
\endfoot
   1 &  PBC J0006.3+2012&                  Mrk 335& Sy1 &   1.58873&  20.20689 &3.10&   0.480&   11.46&$   1.88\pm0.18$&$   0.95\pm0.31 $&0.0254&43.34&3~~1~~n~~h~~a\\
   2 &  PBC J0009.3-0034&  2MASX J00091156-0036551& Sy2 &   2.34193&  -0.57510 &3.89&   3.558&    5.35&$   0.92\pm0.17$&$           ... $&0.0730&44.17&1~~1~~n~~h~~b\\
   3 &  PBC J0010.4+1059&                 Mrk 1501& BLA &   2.61064&  10.99384 &2.59&   1.579&   10.02&$   1.80\pm0.18$&$   0.79\pm0.27 $&0.0893&44.65&3~~1~~n~~h~~a\\
   4 &  PBC J0018.3+8135&               S5 0014+81& BLA &   4.58770&  81.59197 &3.34&   2.682&    6.75&$   1.02\pm0.16$&$   0.62\pm0.39 $&3.3660&48.12&1~~1~~n~~h~~a\\
   5 &  PBC J0023.0+6138&          IGR J00234+6141& CV* &   5.77389&  61.64727 &3.91&   3.795&    7.53&$   1.12\pm0.20$&$   0.20\pm0.27 $&   ...&  ...&2~~1~~y~~l~~a\\
   6 &  PBC J0024.9+6407&                TYCHO SNR& SNR &   6.23328&  64.13138 &3.26&   2.824&    6.99&$   1.14\pm0.18$&$   0.39\pm0.30 $&   ...&  ...&3~~1~~y~~l~~a\\
   7 &  PBC J0025.0+6248&          IGR J00245+6251&  gB &   6.27016&  62.80854 &4.16&   4.688&    4.85&$   0.70\pm0.18$&$   0.16\pm0.38 $&   ...&  ...&1~~1~~y~~l~~a\\
   8 &  PBC J0025.6+6823&          IGR J00256+6821& AG? &   6.40185&  68.39235 &3.47&   1.834&    6.37&$   1.00\pm0.17$&$   0.82\pm0.48 $&0.0120&42.50&1~~1~~n~~h~~a\\
\end{longtable}
\begin{list}{}{}
    \item[$^{\star}$] The source type is coded according to the nomenclature used in SIMBAD.
    \item[$^{\dag}$] Flux is in units of $10^{-11}$\ferg.
    \item[$^{\ddag}$] Flag A: energy band with highest significance (1=14--150
    keV; 2=14--30 keV; 3= 14--70 keV);\\
    Flag B: energy band used for the calculation of the flux  (1=14--150 keV;
    2=14--30 keV; 3= 14--70 keV);\\
    Flag C: y if already reported as hard X-ray source;\\
    Flag D: l if the source has $|b|<5^{\circ}$, h if the source has $|b|>5^{\circ}$;\\
    Flag E: strategy used for the identification (see Sect.\ref{id});\\
\end{list}
\end{landscape}
\clearpage

\end{document}